\documentclass{article}

\usepackage{PRIMEarxiv}

\usepackage[utf8]{inputenc} % allow utf-8 input
\usepackage[T1]{fontenc}    % use 8-bit T1 fonts
\usepackage{hyperref}       % hyperlinks
\usepackage{url}            % simple URL typesetting
\usepackage{booktabs}       % professional-quality tables
\usepackage{amsmath,amssymb}       % blackboard math symbols
\usepackage{nicefrac}       % compact symbols for 1/2, etc.
\usepackage{microtype}      % microtypography
\usepackage{lipsum}
\usepackage{fancyhdr}       % header
\usepackage{graphicx}       % graphics
\graphicspath{{media/}}     % organize your images and other figures under media/ folder
\usepackage{xcolor}
\usepackage{bm}
\newcommand{\rr}{{\bm{r}}}
\newcommand{\md}{{\rm{d}}}

\newcommand{\mi}{{\rm{i}}}
\newcommand{\ms}{ms$^{-1}$}
\newcommand{\dt}{\frac{\partial}{\partial t}}
\usepackage{svg}

\title{Atomic diffraction by nanoholes in hexagonal boron nitride
}

\author{
  Eivind Kristen Osestad\thanks{Lace Lithography AS, Bergen, Norway.} \\
 Department of Physics and Technology\\ University of Bergen \\
  Bergen, Norway \\
   \And
  Ekaterina Zossimova\thanks{Freiburg Center for Interactive Materials and Bioinspired Technologies (FIT), University of Freiburg, Freiburg, Germany.} \\
  Department of Physics and Astronomy\\ Living Systems Institute\\University of Exeter \\
  Exeter, UK\\
  \AND
  Michael Walter\footnotemark[2] \\
  Institute of Physics\\University of Freiburg\\Freiburg, Germany \\
 \And
Bodil Holst\footnotemark[1] \\
 Department of Physics and Technology\\ University of Bergen \\
  Bergen, Norway\\
  \And
Johannes Fiedler \\
 Department of Physics and Technology\\ University of Bergen \\
  Bergen, Norway\\
  \texttt{johannes.fiedler@uib.no} \\
}

\begin{document}
\maketitle

\begin{abstract}
Fabricating patterned nanostructures with matter waves can help to realise new nanophotonic devices. However, due to dispersion effects, designing patterns with nanoscale features is challenging. Here, we consider the propagation of a helium matter wave through different holes in hexagonal boron nitride (h-BN) as a case study for the weakest dispersion interaction and the matter wave's diffraction as it passes through the holes. We use a quantum-mechanical model to calculate the polarisability of edge atoms around the holes, where we observe polarization ripples of enhanced and reduced polarisabilities around the holes. We use these values to calculate van der Waals dispersion coefficients for the scattered helium atoms. We find that the resulting diffraction patterns are affected by the shape and size of the holes, where the smallest holes have a radius of just $6$~\AA. 
These results can be used to predict the resolution limits of nano-hole patterns on nanophotonic materials.
\end{abstract}

% keywords can be removed

\section{Introduction}
\begin{figure}
    \centering
    \includegraphics[width=0.7\columnwidth]{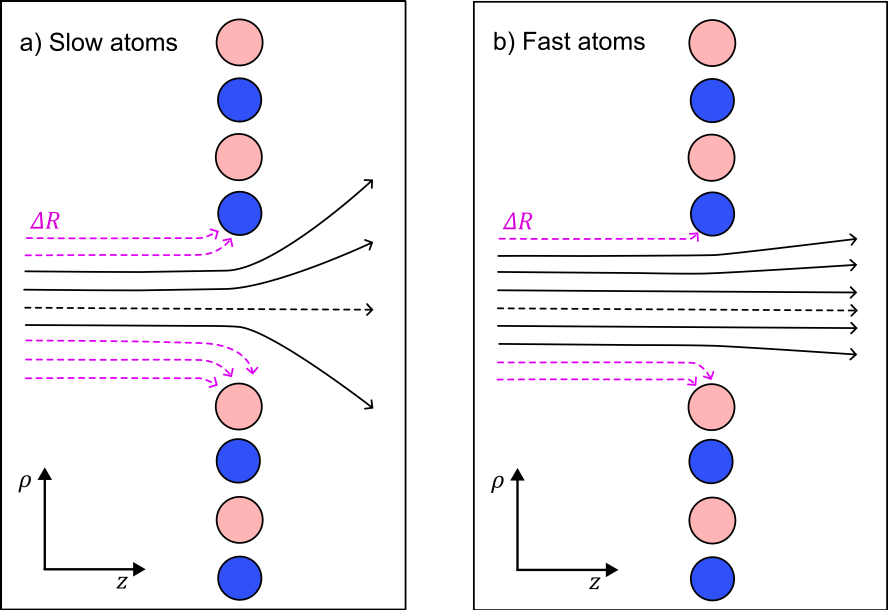}
    \caption{Schematic of the hole reduction $\Delta R$ for (a) slow helium atoms, and (b) fast helium atoms passing through a hole in hexagonal boron nitride (h-BN). The dispersion forces are stronger for slow atoms compared to fast atoms, therefore the hole reduction effect is greater in part (a) compared to part (b).  Additionally, the helium matter wave in part (a) experiences a greater angular spread compared to part (b). The transmission close to nitrogen atoms (blue) increases compared to boron atoms (pink) since nitrogen atoms have a reduced C6 coefficient compared to boron atoms. This leads to interesting features in the phase shift plots and diffraction patterns.}
    \label{fig:hole-reduction}
\end{figure}

Atomic interferometry is useful for precise measurement that can be applied in fundamental physics tests\cite{Tino_2021} and accurate inertial sensing~\cite{PhysRevLett.116.183003}. It is common to use material gratings in such experiments in order to make the atom interfere~\cite{RevModPhys.81.1051}. For very small holes or slow atoms, dispersion forces between the interfering atoms and the mask begin to have a significant effect. Examples of such forces include the van der Waals force between neutral particles; and the Casimir--Polder force between neutral particles and dielectric materials~\cite{Buhmann12a, Brand15}. These forces arise from ground state fluctuations of the electromagnetic fields and reduce the effective size of the holes, as shown in Fig.~\ref{fig:hole-reduction}. Furthermore, they produce a phase shift that affects the atomic waves passing through the holes~\cite{Fiedler_2022}. This limits the size of the holes in a given diffraction mask.

Since the dispersion forces depend on the thickness of the mask, 2D monolayer materials, such as hexagonal boron nitride (h-BN)~\cite{ryu_atomic-scale_2015} and graphene~\cite{Brand2015}, represent the theoretical lower limit to these interactions. Graphene has already been used as a beam splitter for matter waves in experiments by creating lines and holes in it~\cite{Brand2015}. 
In addition, by firing high-speed hydrogen and helium atoms at the membrane (at a speed in excess of $27000$ \ms), the matter wave should semi-coherently diffract through even the hexagonal grid in 2D materials~\cite{Brand_2019}.

Many advanced technologies exploit quantum effects at the nanoscale, requiring highly controlled and precise fabrication techniques.
Examples include quantum electronic devices, such as resonant tunnelling diodes, single-electron Coulomb blockade transistors~\cite{Kastner92, Putnam2017}, and quantum dot transistors~\cite{Zhuang98}. However, fabrication techniques often limit the realisation of new devices. For example, ferromagnetic semiconductors could be used in the next generation of energy-efficient computers and electronic devices, which rely on the quantum control of spin states instead of charge carriers~\cite{Ando06}. However, fabricating ferromagnetic semiconductors is challenging and requires new methods to pattern magnetic materials as sub-nanometre dots.

Currently, it is only possible to pattern arbitrary structures with resolution and pitch on the "few-nanometre" scale using electron or ion beam lithography or a scanning probe tip~\cite{Martínez2007}. 
These techniques all write patterns in a series, one pixel at a time, which makes them unsuitable for large-scale industrial applications as it is too time-consuming to pattern large areas. Thus, the lithography industry is dominated by photolithography due to its much higher speed, despite its lower resolution. The current state of the art is extreme-ultra-violet (EUV) lithography. This uses light with a wavelength of $13.5\, \rm{nm}$, corresponding to an Abbe resolution limit of $6.75\, \rm{nm}$, assuming the maximum value of the numerical aperture (NA$=1$). Beyond the Abbe resolution limit, electron blurring of the pattern from secondary effects in the resist material limits the ultimate resolution to approximately $6\, \rm{nm}$ for any photon-based lithography~\cite{Fan2016}.

Binary holography methods using metastable atoms have been proposed as a solution to the challenge of speeding up and improving the resolution of this process~\cite{Nesse17, Nesse19, fujita1996manipulation}. The early proposals did not account for dispersion forces between atoms and the mask~\cite{Scheel2008, WIPS}.
This is a problem, as when using binary holography, very small holes are needed to create patterns of high resolution~\cite{Nesse17, Nesse19}, and thus these forces must be accounted for. A structure of holes in a membrane, a so-called "atom sieve", has been used to focus neutral helium atoms, analogous to the "photon sieve" solution~\cite{flatabo2017atom} for the purpose of neutral helium atom microscopy~\cite{palau2023neutral}.

In the case of h-BN considered here, additional electrostatic forces may appear near the hole as the single charges do not compensate for each other.
These forces are attractive and have a smaller but similar effect on surrounding atoms compared to the dispersion forces~\cite{2011xix}.

In this paper, we present a model describing the diffraction of a neutral, ground-state helium matter wave through a monolayer. More specifically, we consider helium atoms passing holes in h-BN. We chose these materials as they are both available for experiments, and they both have weak dispersion interactions due to helium being a small, inert atom and h-BN being an insulator, thus the dispersion interaction will be weak. By removing atoms from the lattice structure of the h-BN monolayer, we create four stable holes that vary in size and shape from about $5-15$ Å across. We note that atomically precise holes can be created in real life with electron beams~\cite{Bui2023}. By modelling the atoms quantum mechanically, we find the forces acting on the atoms passing through the hole and the effective reduction in the radii of the holes. Then, we use the hole reduction and the forces to calculate the diffraction patterns macroscopically.  

First, in section~\ref{seq:Modelling}, we use electronic structure theory and dispersion force theory to model defects in h-BN. We find several stable structures corresponding to charge-compensated holes in h-BN. We determine the polarisability and van der Waals coefficients of edge atoms around the defects, which allows us to calculate the corresponding forces around the hole. Then, in section~\ref{seq:Hole_reduction}, we estimate the hole reduction by simulating a helium atom colliding with boron and nitrogen atoms. Having found both the forces and the hole reduction, we switch to a macroscopic diffraction picture in section~\ref{seq:diffraction-pattern}, due to the difficulty of numerically modelling the atomic wavefunction. We calculate the resulting shape of the hole and how the forces inside and outside the hole change the phase of the matter wave passing through it. Using this information, we find a transmission function, telling us where the atom can pass through the hole and the phase change at that point. This can then be used to find the far-field diffraction patterns of the holes.

\section{Modelling}\label{seq:Modelling}

\subsection{Screened atomic polarisabilities}

\begin{figure*}
    \centering
    \includegraphics[width=\textwidth]{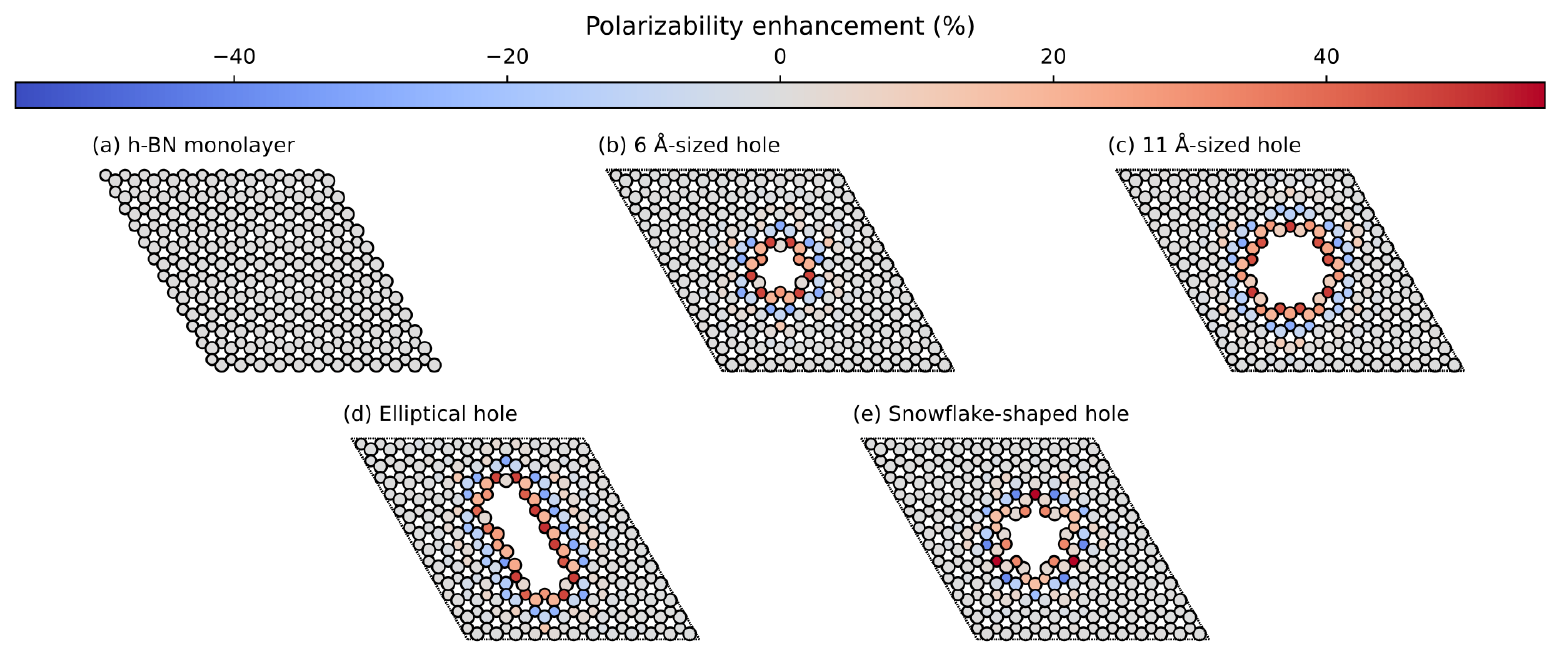}
    \caption{Polarisability ripples around different shaped holes in the h-BN monolayer. The colour bar represents the percentage increase (red) or decrease (blue) of the atomic polarisabilities compared to equivalent atoms in the infinite h-BN monolayer. Geometries visualised with the Atomic Simulation Environment~\cite{larsen_atomic_2017} and overlaid with polarisability data using Matplotlib.~\cite{Hunter:2007}}
    \label{fig:polarisability-enhancement}
\end{figure*}

We obtain the screened atomic polarisabilities from the electronic structure theory.
In the first step, the structures of pristine h-BN and several defects are constructed with the help of the atomic simulation environment~\cite{larsen_atomic_2017}. h-BN has a hexagonal lattice structure, similar to graphene, with a lattice constant $a_{\rm l}= 0.2504\,\rm{nm}$~\cite{C7RA00260B,doi:10.1080/10584587.2015.1039410}, leading to a bond-length between boron and nitrogen of $a= 0.1446 \,\rm{nm}$. We consider only neutral holes where the number of removed B and N atoms is equal in order to avoid the need for compensation charges in periodic calculations. 
The resulting structure is then relaxed until the maximal force on each atom is below $0.05$ eV/\AA. The electronic structure is determined within DFT as implemented in the open-source GPAW code~\cite{mortensen_real-space_2005, Enkovaara2010}. The exchange-correlation energy is described by the PBE functional~\cite{perdew_generalized_1996}.
Kohn-Sham wave functions and electron density are represented in Blöchl's projector augmented wave method~\cite{blochl_projector_1994} and the smooth wave functions are represented on real space grids with a grid spacing of $0.2$\,\AA. 

We follow the Tkatchenko--Scheffler approach~\cite{tkatchenko_accurate_2009, tkatchenko_accurate_2012} to correct the atomic polarisabilities for effects through the constraints of their interaction within the material. Free atomic polarisabilities, $\alpha_i^{\textrm{free}}$, are taken from the Chu and Dalgarno dataset~\cite{Chu2004} and are isotropic. The polarisability of the bounded atoms, $\alpha_i^{\textrm{hirsh}}$, are known to scale approximately linearly~\cite{brinck_polarizability_1993} with the ratios of bonded-atom volumes to free atom volumes
\begin{equation}
    \frac{\alpha_i^{\textrm{hirsh}}[n(\textbf{r})]}{\alpha_i^{\textrm{free}}} \approx \frac{V_i^{\textrm{hirsh}}[n(\textbf{r})]}{V_i^{\textrm{free}}}.
\label{eq:alpha-static}
\end{equation}
The volume ratios, $V_i^{\textrm{hirsh}} / V_i^{\textrm{free}}$, are calculated from the electronic density of the h-BN monolayer, $[n(\textbf{r})]$, using the Hirshfeld charge partitioning scheme~\cite{Hirshfeld1977}.

We subsequently apply a correction for screening between neighbouring atoms using the range-separated self-consistent screening method available in the libMBD code.~\cite{ambrosetti_long-range_2014, bucko_many-body_2016} 
Since the h-BN supercell is periodic, a cut-off radius determines neighbouring atoms, and calculations are truncated according to the Ewald summation method.~\cite{ewald_berechnung_1921, ewald_introduction_1969} The screened atomic polarisabilities are obtained by solving the self-consistent screening equation from classical electrodynamics
\begin{equation}
    \boldsymbol{\alpha}_i = \alpha_i^{\textrm{hirsh}} \left( \textbf{1} - \sum_{i \neq j} \bm{\mathcal{T}}_{ij} \boldsymbol{\alpha}_j \right),
    \label{eq:alpha-screened}
\end{equation}
where $\boldsymbol{\alpha}_i$ denote the screened atomic polarisability tensors and $\alpha_i^{\textrm{hirsh}}$ are the Hirshfeld-partitioned atomic polarisabilities given by Eq.~(\ref{eq:alpha-static}). Here, $\bm{\mathcal{T}}_{ij}$ denotes the dipole-dipole interaction tensor, which depends on the relative displacement of the atoms in the h-BN monolayer. The implementation in the libMBD code attenuates the short-range interactions between atoms to avoid unphysical values at short interatomic separations. 

\subsection{Screened van der Waals coefficients}

Calculating the van der Waals dispersion coefficients requires a model for the dynamic polarisability of atoms. We can introduce frequency dependence through the single-pole approximation
\begin{equation}
    \alpha_i(\mi\xi) = \frac{\alpha_i}{1 + \{\xi / \xi_i\}^{2}},
    \label{eq:alpha-dyn}
\end{equation}
where $\alpha_i=\mathrm{Tr}[\boldsymbol{\alpha}_i]/3$ denotes the scalar, screened polarisability of an atom in the h-BN membrane, evaluated using Eq.~(\ref{eq:alpha-screened}). $\xi_i$ is the corresponding characteristic resonance frequency, approximated by~\cite{tkatchenko_accurate_2009} 
\begin{equation}
    \xi_i = \frac{4(4\pi \varepsilon_0)^2}{3\hbar} \frac{C_{6,\,ii}^{\textrm{free}}}{\left(\alpha_i^{\textrm{free}}\right)^{2}}.
    \label{eq:characteristic-freq}
\end{equation}
using the free atom values for the polarisabilities and van der Waals $C_6$ coefficients.\cite{Chu2004} This represents a suitable approximation for $\xi_i$ because the scaling factors due to Hirshfeld partitioning approximately cancel each other in this ratio. 

The screened van der Waals coefficients can then be obtained using the Casimir--Polder integral, 
\begin{equation}
    C_{6,\,ij} = \frac{3\hbar}{\pi(4\pi \varepsilon_0)^2} \int\limits_0^\infty \md \xi \alpha_i(\mi\xi)\alpha_j(\mi\xi) \,,
    \label{eq:cpolder}
\end{equation}
which can be solved by substituting Eq.~(\ref{eq:alpha-dyn}) into Eq.~(\ref{eq:cpolder}). This leads to the London formula~\cite{tkatchenko_accurate_2009, Tang1969},
\begin{equation}
    C_{6,\,ij} = 
    \frac{3\hbar}{2(4\pi \varepsilon_0)^2} 
    %\frac{3}{2} 
    \frac{\xi_i \xi_j}{\left(\xi_i + \xi_j \right)} \alpha_i \alpha_j \,,
\label{eq:london}
\end{equation}
in terms of the screened atomic polarisabilities $\alpha_i=\mathrm{Tr}[\boldsymbol{\alpha}_i]/3$ and characteristic frequencies $\xi_i$.

\begin{table}[htb]
    \centering
    \caption{Polarisability and (homonuclear) $C_6$ coefficients (in atomic units) of boron and nitrogen atoms in the infinite h-BN monolayer and the free helium atom. 1 Bohr$^3$ = $1.48\times10^{-31}$m$^3$, 1 Ha Bohr$^6$ =  $9.57\times10^{-80}$ J m$^6$,
    1 Ha = $4.36\times10^{-18}$ J.}
    \begin{tabular}{c|c c c}
        Atom & $\alpha/(4\pi \varepsilon_0)$ [Bohr$^3$] & $C_6$ [Ha\,Bohr$^6$] & $\hbar\xi$ [Ha]\\
        B & 18.09 & 75.23 & 0.30 \\
        N & 3.70 & 14.62 & 0.59 \\
        He & 1.38 & 1.42 & 0.99
    \end{tabular}
    \label{tab:alpha-C6}
\end{table}
Here, we are interested in the dispersion interactions between free helium particles and the set of $\{j\}$ atoms in the h-BN membrane. Therefore, we can set $\xi_i = \xi^{\textrm{He}}$ and $\alpha_i = \alpha^{\textrm{He}}$ in Eq.~(\ref{eq:london}), reducing the dimensionality of the problem. This gives
\begin{equation}
    C_{6,\,j} = 
    \frac{3\hbar}{2(4\pi \varepsilon_0)^2} 
\frac{\xi_i \xi^{\textrm{He}}}{\left(\xi_i + \xi^{\textrm{He}} \right)} \alpha^{\textrm{He}} \alpha_j  \,,
\label{kz:c6-coeff}
\end{equation}
where the values for the free helium atom are given in Tab.~\ref{tab:alpha-C6}.
The values of $\alpha_j$ depend on the atomic species and the displacement of the atom relative to the defect in the monolayer.

\subsection{Polarisability ripples}

We solve Eq.~(\ref{eq:alpha-screened}) for the five different supercell geometries shown in Fig.~\ref{fig:polarisability-enhancement}. The structures are charge neutral since we remove equal numbers of boron and nitrogen atoms to create the holes. The infinite h-BN monolayer is used as a reference point to calculate the change in atomic polarisabilities for the other structures with differently shaped holes. The baseline values for the infinite h-BN supercell are summarised in Table~\ref{tab:alpha-C6}. In the case of supercells with holes, we observe a "polarisability ripple" around the hole's circumference. The atoms immediately surrounding the hole have an enhanced polarisability compared to equivalent atoms in the infinite h-BN monolayer. The enhancement is about twice as strong for nitrogen atoms ($\approx 40\%$) compared to boron atoms ($\approx 20\%$). The second ring of atoms surrounding the hole shows a decrease in atomic polarisability, although this is a less pronounced change than for the first ring of atoms. This ripple effect propagates outwards from the hole, and the oscillations rapidly decay until they return to the h-BN baseline values for the infinite monolayer. This effect looks like a "polarisability ripple" in Figs.~\ref{fig:polarisability-enhancement} (b)-(e).  
 
\subsection{Dispersion interactions by h-BN monolayers}

\begin{figure}
    \centering
    \includegraphics[width=0.4\columnwidth]{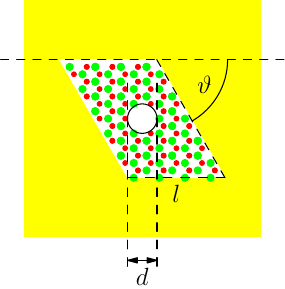}
    \caption{Sketch of a spherical hole with diametre $d$ in an infinite monolayer (yellow area). A small rhombus with side lengths $l$ and angle $\vartheta$ will be treated atomically. The remaining yellow area is a solid object. We always model a large enough area atomically such that we can neglect the solid part.}
    \label{fig:2}
\end{figure}
We now determine the interaction potentials experienced by a helium atom approaching
h-BN monolayers.
We consider an atomically thin monolayer as depicted in Fig.~\ref{fig:2}. A small spherical hole with a diameter $d$ is treated as a vacancy of several atoms. Thus, the interaction is considered on an atomic level in a rhombus with the side lengths $l$ surrounding the hole. The remaining part of the monolayer membrane is treated as a continuum without significant interaction. We particularly consider hexagonal boron nitride (h-BN) and neutral helium atoms. However, the derived models for the interactions and their effective treatments for matter-wave interference can also be adapted easily to other materials.  

The interaction between a neutral particle characterised through its polarisability tensor $\boldsymbol{\alpha}$ and a dielectric object is, in general, given by the Casimir--Polder potential~\cite{Scheel2008,Buhmann12a}
\begin{equation}
    U_{\rm CP}(\rr) = \frac{\hbar\mu_0}{2\pi}\int\limits_0^\infty \md \xi\, \xi^2\operatorname{Tr}\left[\boldsymbol{\alpha}(\mi\xi)\cdot {\bf{G}}(\rr,\rr,\mi\xi)\right]\,,\label{eq:CP}
\end{equation}
with the reduced Planck constant $\hbar$, the vacuum permeability $\mu_0$ and the scattering Green function ${\bf{G}}$ that contains the properties of the dielectric object. This equation can be understood as an exchange of virtual photons with frequency $\mi\xi$, which are induced at the particle's position $\rr$ and back-scattered from the surrounding dielectric objects as described by ${\bf{G}}(\rr,\rr, i\xi)$. The scattered virtual photons polarise the particle. The sum (respectively the integral) over all possible photon exchanges yields the Casimir--Polder interaction.  

Here, for a monolayer membrane, the question of the scattering properties of the membrane arises. In previous works, we derived an approximation for the dispersion interaction for weakly responding materials by integrating over the volume of the dielectric object~\cite{Fiedler_2022}.

Due to this volume integral, the interaction will strongly depend on the assumed thickness of the monolayer, which is not precisely quantifiable for monolayers. For h-BN, the thickness of a monolayer can be approximated by $0.3 \, \rm{nm}$~\cite{doi:10.1021/acs.nanolett.0c04222}. However, as we are interested in the behaviour at very short separations, the thickness's uncertainty will substantially impact the results. 

To avoid issues concerning the monolayer's thickness, we separate the monolayer's surface into sections according to Fig.~\ref{fig:2} with a rhomboidally shaped section (region I) surrounding the spherical hole with diameter $d$. According to the lattice structure, this section is defined by a side length $l$ and a wedge angle $\vartheta$. We used $\vartheta=60^\circ$ for h-BN. 
This section will be treated atomically, whereas the remaining outer part (yellow area; region II) is a non-contributing continuum by sufficiently increasing the size of the region I. This is motivated by the $r^{-6}$ power law for the distance dependence of the short-range dispersive pairwise interaction potential. 
In region I, we consider an atomistic representation of the membrane characterised by the atom's positions $\rr_i$ and its type expressed by a local polarisability $\alpha_i$. This approach yields the discrete form of the first-order of the Born series expansion~\cite{WIPS}
\begin{equation}
{\bf{G}}({\bm{r}},{\bm{r}}',\omega) = \frac{\omega^2}{c^2\varepsilon_0}\sum_i{\bf{G}}({\bm{r}},{\bm{r}}_{i},\omega)\cdot \boldsymbol{\alpha}_i(\omega) \cdot {\bf{G}}({\bm{r}}_i,{\bm{r}}',\omega)\,. \label{eq:Born}
\end{equation}
By plugging Eq.~(\ref{eq:Born}) into the Casimir--Polder potential~(\ref{eq:CP}), the interaction in the region I can be written as the sum over the screened van der Waals interactions
\begin{equation}
    U_{\rm CP} (\rr) = -\sum_j \frac{C_{6,\,j}}{\left|\rr_j -\rr\right|^6}\,, \label{eq:ucpapp}
\end{equation}
with the screened van der Waals coefficient 
from Eq.~(\ref{kz:c6-coeff}).

Equation~\eqref{eq:ucpapp} can be used as a criterion for the width of the region I. By considering a linear atomic chain with period $a_{\rm C}$ in a one-dimensional configuration, the total van der Waals potential for a particle at distance $r$ to the chain is determined by
\begin{equation}
    U_{\rm vdW}(r) =-\frac{C_6}{r^6}\sum_{j=0}^N \frac{1}{\left(1+j a_{\rm C}/r\right)^6}\,.
    \label{eq:chain-of-atoms}
\end{equation}
By restricting the chain to a finite particle number $N$, the deviation between the truncated and infinite sum can be obtained as
\begin{equation}
    \frac{\left|U(r) -\lim_{N\mapsto\infty} U(r)\right|}{\lim_{N\mapsto\infty} U(r)}=\frac{{\rm{Li}}_5\left(N+1+r/a_{\rm C}\right)}{{\rm{Li}}_5\left(r/a_{\rm C}\right)}\,,
    \label{eq:chain-of-atoms-solution}
\end{equation}
with the polylogarithm ${\rm{Li}}_5(x) = \sum_{k=0}^\infty x^k/k^5$. Thus, the error according to the chain length $Na_{\rm C}$ can be estimated, leading to two atoms ($N=2$) for an error below $1$\%. Consequently, two atomic rings surrounding the hole cause almost the entire interaction. For the holes considered here, all layers after the 3rd contribute to less than $1$\% of the total potential inside the holes. 

A directional dependence appears assuming that that $\boldsymbol{\alpha}_i(\omega) = \alpha_i(\omega) \boldsymbol{D}_i$~\cite{D0CP02863K,doi:10.1021/acs.jpca.8b01989,https://doi.org/10.1002/andp.201500224}, with $\boldsymbol{D}_i$ being a $3\times3$-matrix, the Casimir--Polder potential can be written as
\begin{equation}
    U_{\rm CP}(\bm{r}) = -\sum_i \frac{C_6^{(i)}}{6|\bm{r}_i - \bm{r}|^6}\left[{\rm Tr} \boldsymbol{D}_i + 3\frac{ (\bm{r}_i - \bm{r}) \cdot\boldsymbol{D}_i\cdot (\bm{r}_i - \bm{r})}{|\bm{r}_i - \bm{r}|^2}\right] \,.
\end{equation}
These potentials inside hole (b) from Fig. \ref{fig:2} are shown in Fig \ref{fig:potentials}.

\subsection{Electrostatic forces}
\begin{figure}[]
    \centering  
    \includegraphics[width=0.7\columnwidth]{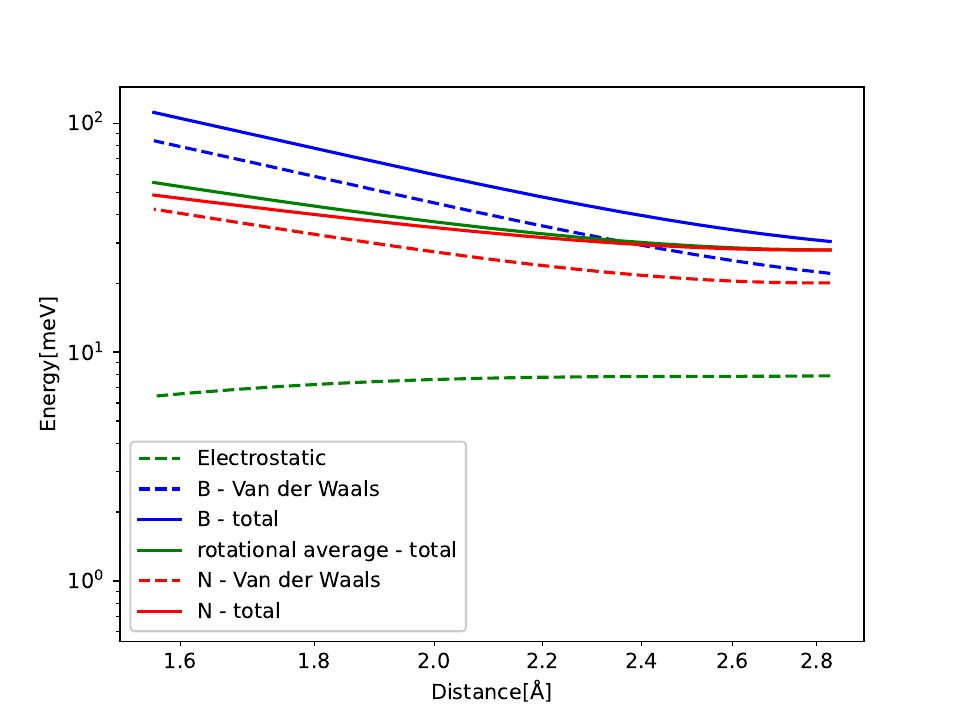}
    \caption{LogLog plot showing the potentials inside the $6$~\AA~hole from Fig.~\ref{fig:polarisability-enhancement}. The lines show the potential from the holes' edges where boron and nitrogen atoms exist. It also shows the different contributions from electrostatic and van der Waals potentials.}
    \label{fig:potentials}
\end{figure}

Such membranes are usually electrical neutral as a bulk system. However, due to the removed atoms creating the hole, the single charges do not compensate each other near the hole. For this reason, each atom at the position $\rr_i$ also carries a charge $q_i$, leading to an induced interaction~\cite{2011xix,https://doi.org/10.1002/prop.201600025}
\begin{equation}
    U_{\rm el} (\rr) = - \frac{\alpha(0)}{2(4\pi\varepsilon_0)^2}\left( \sum_i \frac{q_i }{\left|\rr-\rr_i\right|^2} \right)^2\,,
    \label{eq:Uel}
\end{equation}
with the static polarisability of the free 
helium atom $\alpha(0)$. The charge $q_i$ is 
obtained from Hirshfeld-partitioning~\cite{Hirshfeld1977} of the electron density of h-BN. This leads to charges of $+0.2$|e| on B and $-0.2$|e| on N 
within pristine h-BN in good agreement with previous studies~\cite{altintas_intercalation_2011} (there are severe disagreements for Bader~\cite{bader_atoms_1994,tang_grid-based_2009} charges in the literature~\cite{wang_local_2016,yin_triangle_2010,suzuki_designing_2022} as
shown in ESI$^\dag$). 
However, the atoms near the hole show slightly increased local charges~\cite{yin_triangle_2010}, never exceeding $\pm 0.39$|e|.
The different potentials experienced by the helium atom inside the $6$~\AA~hole from Fig.~\ref{fig:polarisability-enhancement} (b) are shown in Fig.~\ref{fig:potentials}.
We can see that the van der Waals potential is considerably stronger. We will nevertheless also consider the electrostatic contributions in what follows.

\section{Diffraction}\label{seq:Diffraction}
We utilise Kirchhoff diffraction to find the resulting diffraction patterns. We need to find the hole reduction and phase shift caused by the van der Waals and electrostatic forces to do this. We find this by simulating a helium wavepacket passing a boron and a nitrogen atom and seeing how much of the wavepacket passes within the van der Waals radius. This then approximates the hole reduction caused by the surrounding atoms. Afterwards, we find the phase shift caused by the forces using an eikonal approximation as the atoms move very fast. Having both the phase shift and the hole reduction, we define a transmission function and find the diffraction patterns for the investigated holes.

\subsection{Hole reduction}\label{seq:Hole_reduction}
A known effect on atoms passing holes is a hole size reduction due to the forces on the atom attracting it to the edges~\cite{Fiedler_2022, bouton2023quantum}. A previous method used to estimate the hole reduction is to track the classical trajectories of atoms and see how far away they have to pass from the wall to avoid collisions~\cite{Fiedler_2022}. Another method uses numerical simulations of the wavefront to find the diffraction at gratings~\cite{bouton2023quantum}. We will compare the classical trajectory method with numerically solving the propagation of the helium wavefunction colliding with an atom. We will simulate the collision with boron and nitrogen separately as it is very difficult to solve the propagation through the entire hole numerically.
In addition, the forces become much weaker as you move away from the atom such that the vast majority of the force is caused by the closest atom in the monolayer, as demonstrated in 
Eqs.~(\ref{eq:chain-of-atoms}) and (\ref{eq:chain-of-atoms-solution}). 

In both the classical and quantum mechanical 
approaches, we assume the extent of the atom to be equal to its van der Waals radius, which is $1.92$~\AA~for B~\cite{Mantina2009}, and $1.55$~\AA~for N~\cite{Bondi1964}. 
The h-BN layer is extended in $x,y$ direction at $z=0$.
The classical hole reduction, $\Delta R_{\rm classical}$, is then estimated by starting with the initial conditions $\rr = (x, y, z)$, $\dot{\rr} = (0, 0, v)$, where $z = 100$~\AA~is the starting distance from the h-BN plane and $v$ is the initial velocity of the He atom. 
Then we let $\rr$ evolve according to
\begin{equation}
    m\ddot{\rr} = -\nabla U(\rr)\, ,
\end{equation}
with, $m$ the mass of the helium. The potential
\begin{equation}
    U(\rr) = U_{\rm el}(\rr) + U_{\rm vdW}(\rr)
\end{equation}
consists of the electrostatic potential (\ref{eq:Uel}) reducing to
\begin{equation}
    U_{\rm el}(\rr) = -\frac{1}{(4\pi\varepsilon_0)^2} \frac{q_j^2 \alpha(0)}{2 \left|\rr\right|^4} \, ,
    \label{eq:kz:electrostatic}
\end{equation}
for a single atom $j$, 
with the static polarisability of the helium atom $\alpha(0)$
and the van der Waals potential 
from Eq.~(\ref{eq:ucpapp}) for a single atom $j$
\begin{equation}
    U_{\rm vdW}(\rr) = - \frac{C_{6,\, j}}{\left|\rr\right|^6} \, .
\label{eq:kz:vdw}
\end{equation}
The atom propagates in the $z$ direction, perpendicular to the monolayer. We test propagate with several different initial starting positions in the $x,y$ coordinates. The hole reduction, $\Delta R_{\rm classical}$, is then the smallest initial distance in the value of $x,y$ plane where the atom does not pass within the van der Waals radius of the atom.  The values of $\Delta R_{\rm classical}$ are given for several velocities in Table~\ref{tab:hole_reduction}.

We also estimate the quantum mechanical hole reduction, $\Delta R_{\rm quantum}$, by considering the case of a helium wave packet colliding with 
boron or nitrogen atoms.
The  wave packet, $\psi$, evolves according to the Schrödinger equation
\begin{equation}
    -\frac{\hbar^2}{2m} \nabla^2 \psi + U(\bm{r}) \psi = i\hbar\frac{\partial}{\partial t} \psi\, .
\label{eq:schro}
\end{equation}

\begin{table}[htb]
    \centering
    \caption{The hole reduction for helium passing by boron and nitrogen at different velocities. Both the classical results $\Delta R_{\rm classical}$ and the quantum mechanical results $\Delta R_{\rm quantum}$ are given.}
    \begin{tabular}{c|c c c}
        Atom & He velocity[\ms] & $\Delta R_{\rm classical}$[Å]& $\Delta R_{\rm quantum}$[Å]\\
        B & $200$ & 6.2 & 8.1\\
        B & $2000$ & 2.5 & 3.6 \\
        B & $20000$ & 1.9 & 2.3\\
        N & $200$ & 5.9 & 7.8\\
        N & $2000$ & 2.4 & 3.2\\
        N & $20000$ & 1.6 & 1.9\\
    \end{tabular}
    
    \label{tab:hole_reduction}
\end{table}
We used a finite difference scheme to evolve a wave packet by colliding with the atoms. The details of this scheme are detailed in the supplementary information.\dag \cite{ParallelMatrixAlgorithm, niyogi2006introduction, Galiffi2015, Fiedler_2022} 
Starting with a Gaussian wave packet of widths, $\sigma_r = \sigma_z = 8$~Å, 
representing a fifth of the box radius such that negligible amounts of the wave function are at the boundary of the simulation box.
We move the potential at a given velocity $v$ towards it from a distance of $60$~Å to a distance of $40$~Å past the helium wave packet. Finally, we assume that the parts of the wave packet that come within the van der Waals radius of the atom have collided with it. These parts of the wave packet might scatter, lose energy or otherwise lose coherence. We model this loss of coherence as an absorption of this section of the wave packet. The radius of the hole reduction then corresponds to the radius of the sphere that would absorb the same amount without any Van der Waals or electrostatic interactions. We determine the hole reduction using the norm of the wave packet
\begin{equation}
    N_f = \int\limits_0^{\infty} \int\limits_{-\infty}^\infty |\psi_f(r, z)|^2 2\pi r \, dr \, dz \,,
\end{equation}
at the end of the propagation.
We assume that the hole reduction corresponds to the radius of a moving sphere absorbing the part of the wave packet that comes within it such that
\begin{equation}
    \int\limits_{-\infty}^{\infty} \int\limits_0^{\Delta R_{\rm quantum}} \frac{2\pi r}{\sqrt{2\pi^3} \sigma_z \sigma_r^2} \exp{\left[-\frac{r^2}{2\sigma_r} - \frac{z^2}{2\sigma_z}\right]} \, dr \, dz = 1 - N_f\, ,
    \label{eq:R_QM}
\end{equation}
with $\sigma_r$ and $\sigma_z$ being the $r$ and $z$ spread of the wave packet.
Solving Eq.~(\ref{eq:R_QM}) for $\Delta R_{\rm quantum}$ then gives
\begin{equation}
    \Delta R_{\rm quantum} = \sqrt{-\ln (N_f) 2\sigma_r^2}\,,
\end{equation}
with the resulting hole reductions given in Tab.~\ref{tab:hole_reduction}.
In all cases, the quantum approach gives a larger hole reduction radius than the classical trajectory approach. The difference varies between a $19-44$\% increase in the reduction radius.
For a small velocity of $200$ \ms, the hole reduction is so large that the holes in Fig.~\ref{fig:polarisability-enhancement} are completely closed 
for the helium atom, as the hole reduction is greater than the radius of the hole.
Even for velocities of $2000$ \ms, the elliptic hole and the $6$\,Å hole, \ref{fig:polarisability-enhancement} (b) and (d) in Fig.~\ref{fig:polarisability-enhancement}, are too narrow for transmission of helium atoms. Only at velocities of $20000$ \ms, all holes allow for some transmission. The resulting effective shape of the holes can be seen in Fig. \ref{fig:phase-shift}. 

\subsection{Phase shift}
By bypassing a dielectric obstacle, a matter wave experiences a spatial-dependent phase shift due to the interactions between both objects.~\cite{Fiedler_2022,Brand15,Hemmrich16,Gack20} This phase-shift reads
\begin{equation}
    \varphi({\bm{\varrho}}) \approx - \frac{m\lambda_{\rm dB}}{2\pi\hbar^2} \int U({\bm{\varrho}},z) \md z\,,
    \label{eq:phase-general}
\end{equation}
in eikonal approximation, where ${\bm{\varrho}}=(x,y)^T$ are the in-plane coordinates, and $U({\bm{\varrho}},z)$ is the potential experienced by the helium atom. $z$ denotes the direction of the moving particles, and $\lambda_{\rm dB} = h/mv$ is the de Broglie wavelength. This approach means that the particles almost pass the obstacle in straight lines. Thus, the phase can be separated into three contributions: an electrostatic part
\begin{equation}
    \varphi_{\rm el}({\bm{\varrho}}) = \frac{m\lambda_{\rm dB} \alpha(0)}{8\pi \hbar^2 \varepsilon_0} \sum_i \sum_j \frac{q_i q_j}{|\rho-\rr_i|^2 |\rho -\rr_j| + |\rho -\rr_i| |\rho -\rr_j|^2}
    \label{eq:phase-el}
\end{equation}
 and van der Waals part
\begin{equation}
    \varphi_{\rm vdW}({\bm{\varrho}}) = \frac{m\lambda_{\rm dB}}{64\hbar^2}\sum_i C_6^{(i)} \left[\frac{2{\rm Tr} \boldsymbol{D}_i + 3\boldsymbol{D}_{izz}}{\left|{\bm{\varrho}}-\rr_i\right|^5} + \frac{5(\bm{\varrho}-\rr_i)\cdot\boldsymbol{D}_i\cdot({\bm{\varrho}}-\rr_i)}{\left|{\bm{\varrho}}-\rr_i\right|^7}\right] \, ,
    \label{eq:phase-vdW}
\end{equation}
for the interaction with the atomic representation in the region I.

\begin{figure*}
    \centering
    \includegraphics[width=\textwidth]{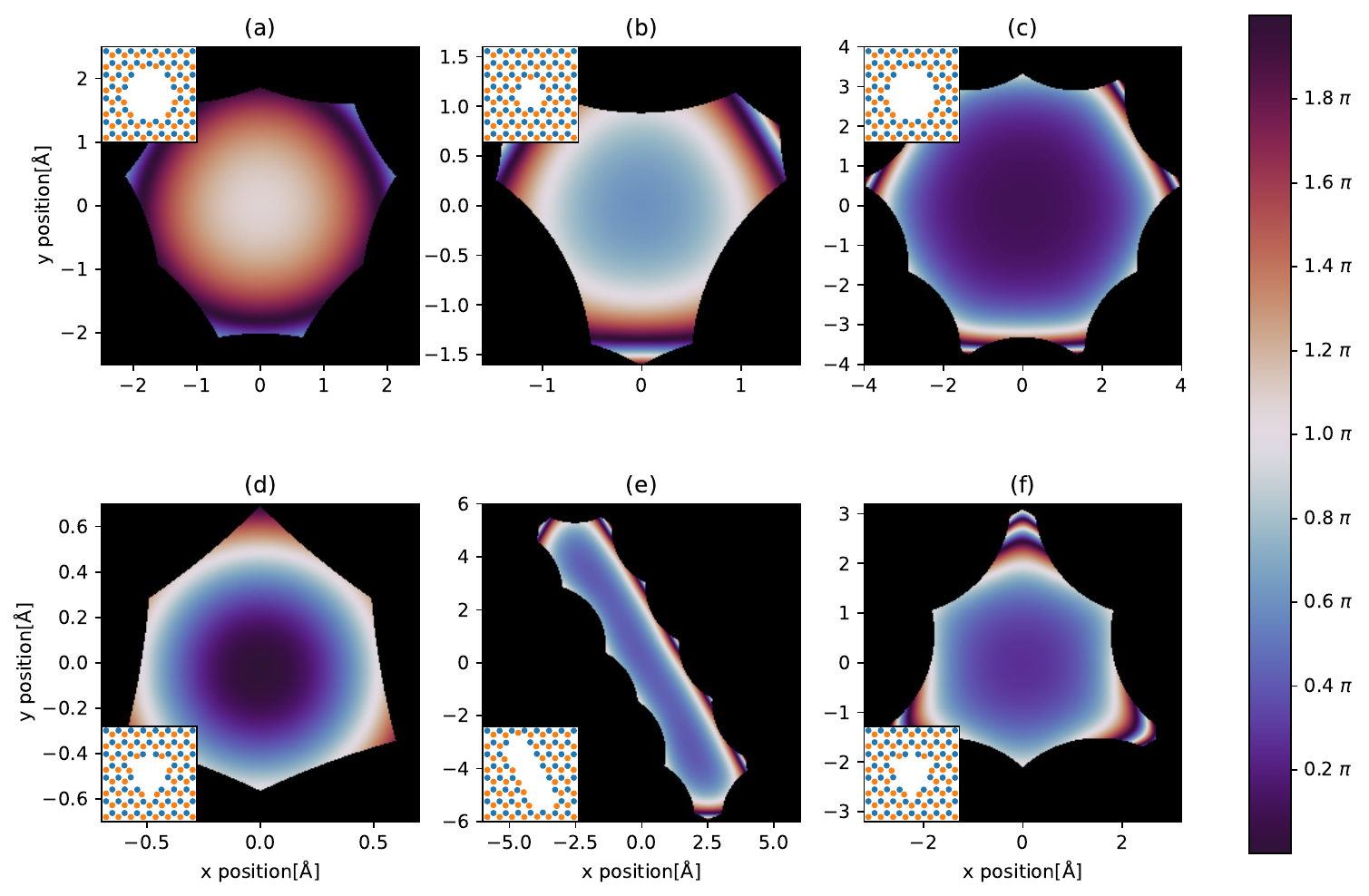}
    \caption{The phase shifts of the helium atoms from interactions with the atoms around the holes shown in Fig.~\ref{fig:polarisability-enhancement}. Black areas mean that the helium atom is not transmitted through the hole. The velocity of the matter wave in parts (a) and (d) is $2000$~\ms, whereas the velocity in parts (b), (c), (e) and (f) is $20000$~\ms. The insets show the layout of the atoms around the hole with B in orange and N in blue. Due to the matter waves interaction with the monolayer, there is an additional reduction in transmission area of (a) $37$\%, (b) $31$\%, (c) $17$\%, (d) $88$\% (e) $5$\% and (f) $15$\% compared to the area of the monolayer covered by the removed atoms.
    }
    \label{fig:phase-shift}
\end{figure*}

We solve these equations to find the phase shift of a matter wave propagating through different types of holes in h-BN. The resulting phase shifts are plotted in Fig.~\ref{fig:phase-shift}, using a cyclical colour map. In parts (b), (c), (e) and (f), the He atoms have a high velocity ($v = 20000$~\ms), whereas in parts (a) and (d) the He atoms have a comparatively low velocity ($v=2000$~\ms). The de Broglie wavelength, $\lambda_{\rm dB}$, is inversely proportional to the velocity. Therefore, from Eq.~\ref{eq:phase-general}, we expect the phase shift to be smaller when the velocity of He atoms is higher (shorter $\lambda_{\rm dB}$). 

This effect can be seen in Fig.~\ref{fig:phase-shift}, where the blue regions at the centre of the holes in parts (b), (c), (e) and (f) correspond to a low phase shift. The phase shift through the central region in part (b) is larger than in part (c) due to the smaller size of the hole ($6$~\AA~vs $11$~\AA).
Therefore, the matter wave experiences stronger dispersion interactions with the edge atoms. By comparison, there is a much larger phase shift at the centre of the holes in parts (a) and (d) due to the slower velocity (longer $\lambda_{\rm dB}$) of the matter wave. We can directly compare parts (a)/(c) and (d)/(f) to see this effect since the potentials around these pairs of holes are the same.

We also observe the h-BN lattice structure's effect in all the phase shift plots. Boron atoms have a higher dispersion coefficient compared to nitrogen atoms. Therefore, in all cases, we observe greater transmission and more fringes near nitrogen-terminated edges. For example, this can be clearly seen in the elliptical hole in part (e), where the nitrogen-terminated edge (long right edge) has features that are not observed on the boron-terminated edge (left long edge). 

Another example is the snowflake structure in part (f), which has unequal transmission through the different arms of the snowflake. The structure has six arms, whereas only three arms strongly transmit matter waves. Each arm has three nitrogen edge atoms or three boron edge atoms (these alternate for each arm around the hole). We observe a high transmission in the arms where nitrogen atoms dominate, along with rapidly oscillating phase shift patterns. It could be interesting to compare this result with an atomically homogeneous structure, such as graphene, where we would expect to see a 6-fold, rather than 3-fold, rotational symmetry in the phase shift pattern. 

By comparison, in part (d), we do not observe the rapidly oscillating features present in part (f). We attribute this to the lower velocity of the matter wave in part (d) compared to (f), which leads to stronger dispersion interactions and reduced transmission through the arms of the snowflake.

\subsection{Diffraction Patterns}\label{seq:diffraction-pattern}

From the phase shifts and the hole reduction, we can derive a transmission function $T(\bm{\rho})$, which is $e^{i\varphi(\bm{\rho})}$ wherever the helium atom is transmitted through the hole and zero everywhere else. 
The diffraction pattern from passing a hole is given by Kirchhoff’s diffraction formula~\cite{BornWolfOptics}
\begin{equation}
    \Psi(\bm{r}) = \frac{A k}{2\pi i}\int d^2 \rho T(\bm{\rho}) \frac{e^{i k (r_0 + s)}}{2 r_0s}\left[ \cos(n, r_0)-\cos(n, s)\right]\,.
\end{equation} 
Here, $A$ is the amplitude of the wave function, and $k = 2\pi / \lambda_{\rm dB}$. $r_0$ is the distance between the point source of the helium matter wave and the in-plane coordinates $\bm{\rho}$ of the h-BN monolayer. $s$ is the distance between $\bm{\rho}$ and the coordinates $\bm{r}$ at which the diffraction pattern is measured $(s=|\bm{r} - \bm{\rho}|)$. The sum of $r_0$ and $s$ describes the total path length of the helium matter wave. $\cos(n,r_0)$ and $\cos(n,s)$ describe the cosine similarity between $r_0$, $s$ and the normal $n$ of the h-BN plane. The cosine similarity is defined as
\begin{equation}
    \cos(n,s) := \cos(\theta) = \frac{\bm{n} \cdot (\bm{r} - \bm{\rho})}{ns}
\end{equation}
with an equivalent expression for $\cos(n, r_0)$.

\begin{figure*}
    \centering
    \includegraphics[width=\textwidth]{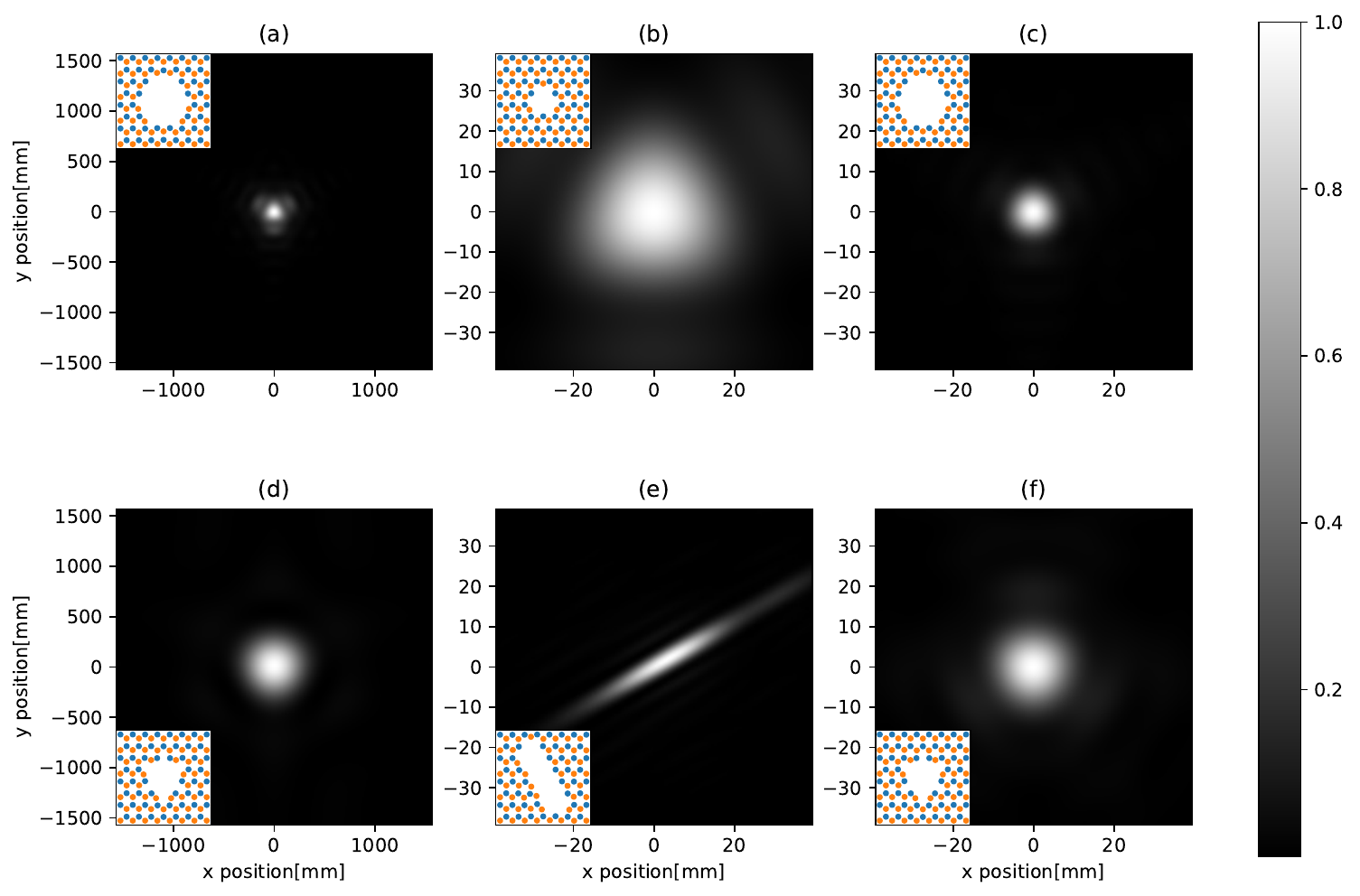}
    \caption{Diffraction pattern resulting from a helium atom passing the holes in Fig.~\ref{fig:polarisability-enhancement} normalised to the maximum intensity. The patterns are 1m away from the holes. Fig (a) and (d) have a velocity of $2000$~\ms, and (b), (c), (e) and (f) have a velocity of $20000$~\ms. The insets show the layout of the atoms around the hole with B in orange and N in blue}
    \label{fig:diff-pattern}
\end{figure*}

If the path length of the matter wave is large compared to the linear dimensions of the aperture, we can apply the small angle approximation, whereby $\left[ \cos(n, r_0)-\cos(n, s)\right] \rightarrow 2 \cos(\delta)$. Here, $\delta$ is the angle between the helium matter wave's displacement vector and the h-BN plane's normal vector. We can further assume that the distances $r_0$ and $s$ are measured from the origin of the coordinate system rather than an arbitrary point in the h-BN plane. These transformed distances are $r_0 \rightarrow r_0'$ and $s \rightarrow |\bm{r}|$. This leads to the Fraunhofer approximation~\cite{BornWolfOptics}, 
\begin{equation}
    \Psi(\bm{r}) = \frac{A k}{2\pi i} \frac{e^{i k (r_0' + r)}}{r_0' r}\int d^2 \rho T(\bm{\rho}) e^{ik \bm{\rho} \cdot \bm{k}}\, ,
\end{equation}
where we have assumed that $\cos(\delta) \approx 1$ for normal incidence. The diffraction pattern at a screen is then given by the absolute value squared of $\psi(\bm{r})$
\begin{equation}
    |\Psi(\bm{r})|^2 = \left|\frac{A k}{2\pi i} \frac{e^{i k (r_0' + r)}}{r_0' r}\int d^2 \rho T(\bm{\rho}) e^{ik \bm{\rho} \cdot \bm{k}}\right|^2\, .
    \label{eq:diff-pattern}
\end{equation}
The resulting diffraction patterns, normalised such that the maximum value is one, are shown in Fig.~\ref{fig:diff-pattern}. When the velocity of the helium atom is $2000$~\ms, the angular spread is very large so it should be noted that the Fraunhofer approximation holds less well.

Figure~\ref{fig:diff-pattern} shows that the h-BN lattice structure directly affects the diffraction patterns. In all cases, by comparing the patterns from Fig.~\ref{fig:phase-shift}, we see more fringes close to nitrogen atoms, which have a lower dispersion coefficient than boron atoms. This is clearly visible in parts (a), (b), (c) and (f), where we observe a 3-fold, rather than 6-fold rotational symmetry in the phase-shift plot.

Part (d) has a much more uniform phase shift across the hole; thus, the diffraction pattern is more similar to how light would diffract through a similarly shaped hole. In part (f), which is the same hole only diffracted by faster atoms, we see a much clearer fringe pattern as we also have diffraction through some of the arms of the snowflake where there are atoms on almost all sides.

\begin{figure}
    \centering
    \includegraphics[width=0.7\columnwidth]{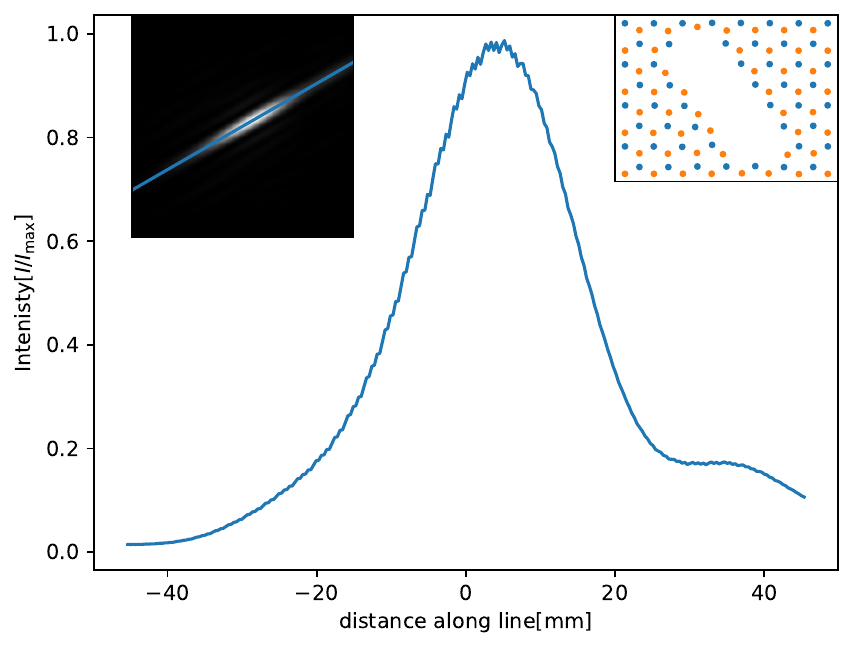}
    \caption{Diffraction pattern along the indicated line for the elliptic hole (Fig. \ref{fig:diff-pattern}(e)). We clearly see more atoms go to the right side than the left. The centre of the pattern is also shifted to the right. The insets show the line we plot along and the layout of the atoms around the hole, with B in orange and N in blue.}
    \label{fig:elliptic-pattern}
\end{figure}
For narrow slits, such as in part (e), we would expect the diffraction pattern to spread more along the short axis of the hole compared to the long axis. While we see this, we can also see that the diffraction pattern is stretched along the right side of the short axis.

Even though the nitrogen atoms have a larger polarisability enhancement, they still have a smaller dispersion coefficient compared to boron atoms. Therefore, there is some asymmetry in the transmission function, leading to a diffraction pattern that is not centred directly in the middle of the elliptical hole but that is shifted closer to the nitrogen-terminated edge. Essentially, the helium atoms are gaining angular momentum from one side having a stronger potential and being "shot" right. This is illustrated in Fig.~\ref{fig:elliptic-pattern}.

\section{Conclusions}
This paper describes a method of determining diffraction patterns from matter waves passing through holes in monolayer materials, using h-BN as an example.
Our DFT calculations reveal
that the removal of atoms to create the hole produces a
"polarisability ripple" around the defect where the atoms surrounding the defect show enhanced polarisability,
whereas the atoms in the next ring show reduced polarisability. 
This oscillatory behaviour diminishes rapidly with increasing distance to the defect.

Based on these polarisabilities we have estimated the hole reduction and phase shift resulting from van der Waals and electrostatic interactions between the atom and the h-BN monolayer through a numerical wave packet propagation considering the edge atom nearest to the classical trajectory of the helium scattering path. We find the van der Waals contribution to dominate 
the scattering potential as compared to the electrostatic part.
Using macroscopic diffraction theory, the propagation allowed us to find diffraction patterns of holes smaller than $1~\rm{nm}$ in h-BN. 
We found that the predicted atomic polarisabilities and dispersion coefficients significantly affect the diffraction patterns, such as shifting the elliptical holes' diffraction pattern in the direction of the nitrogen-terminated edge. Our results suggest that matter-wave lithography can achieve sub-nanometer resolution by using 2D monolayer materials as a mask. %\kz{Our results suggest that matter-wave lithography can / cannot achieve sub-nanometer resolution by using 2D monolayer materials as a mask? (the transmission area is Angstrom sized from the phase shift plots but the diffraction patterns have a diameter of approx. 4cm in parts (b), (f) - does this mean that we don't see sub-nanometer resolution due to strong dispersion interactions?)}\ek{The diffraction pattern does not matter as the phase still changes rapidly, so interference between multiple holes can allow for subnanometer resolution.}

\section*{Acknowledgements}
J.F. gratefully acknowledges support from the European Union (H2020-MSCA-IF-2020, grant number: 101031712).
B.H., E.K.O. and J.F. gratefully acknowledge support from the European Union (H2020-FETOPEN-2018-2019-2020-01, grant number: 863127 Nanolace).
E.Z. and M.W. acknowledge support from the German Research Foundation 
(grant number: WA 1687/10-1); 
E.Z. and M.W. are thankful for the computing resources provided by the state of Baden-Württemberg through bwHPC and the German Research Foundation through grant number INST 40/575-1 FUGG (NEMO and JUSTUS 2 clusters).

%Bibliography
\bibliographystyle{unsrt}  
\bibliography{references}  

\newpage
\section{Supplemental Information}

\subsection{Wave propagation simulation}
We want to estimate how much the effective size of our holes shrink due to the dispersion forces as this is a know effect in other holes\cite{Fiedler_2022}. We used two methods to do this. Here we give a more detailed description of the second method which is a finite difference Cranck-Nicholson scheme. 

To estimate the reduced radius of the holes we will simulate a helium wave-packet colliding with the boron and nitrogen atoms. We will assume that any part of the wave packet that comes within the Van der Waals radius of the boron and nitrogen is absorbed and disappears. Outside the VdW radius, we assume that the atom obeys the Schrödinger equation with a Van der Waals potential and an electrostatic potential
\begin{equation}
    \frac{\hbar^2}{2m} \nabla^2 \psi(\bm{r}, t) - \frac{C_6}{|\bm{r} - \bm{r}_A|^6}\psi(\bm{r}, t) - \frac{1}{4\pi \varepsilon_0}\frac{q^2 \alpha(0)}{|\bm{r}-\bm{r}_A|^4}\psi(\bm{r}, t) = i\hbar \dt \psi(\bm{r}, t) \, ,
\end{equation}
with $\bm{r}_A$ being the position of the nitrogen or boron atom, $\bm{r}$ being the position in the simualted box, 

The general form of the Schrödinger equation is 
\begin{equation}
    \hat{H}\psi = i \hbar \dt \psi \, ,
\end{equation}
which has a solution on the form 
\begin{equation}
    \psi(\bm{r}, t) =  \exp(-\frac{i\hat{H}}{\hbar}(t-t_0)) \psi(\bm{r}, t_0)\, .
\end{equation}
A half implicit half explicit finite difference method is then used to approximate the propagation
\begin{equation}
    \label{eq:timestep_long}
    (1 + \frac{i \hat{H}}{2\hbar}\delta t) \psi_{t+\delta t} =(1 - \frac{i \hat{H}}{2\hbar} \delta t) \psi_t \, .
\end{equation}
This finite difference scheme is the same as used by \cite{Galiffi2015}, we use it as it is norm preserving.
Now we define
\begin{equation}
    (1 \pm \frac{i \hat{H}}{2\hbar}\delta t) = \mathcal{A}_\pm\, ,
\end{equation}
such that equation \ref{eq:timestep_long} becomes
\begin{equation}
    \mathcal{A}_+ \psi_{t+\delta t}= \mathcal{A}_-  \psi_t\, .
\end{equation}

To reduce the number of points having to be simulated we assume the helium atom is stationary and that the potential is moving towards it. In addition, we assume that the collision is head-on and that there is no angular momentum relative to the helium atom. Thus we can model a box around the helium atom at rest and move the boron or nitrogen into the box during the simulation. We can also assume that the wavefunction can be split into an angular part which we assume is uniform and a radial and z-component part which evolves in time. The resulting Schrödinger equation then looks like
\begin{equation}
\label{eq:schr_long}
    \frac{\hbar^2}{2m}(\frac{\partial^2}{\partial z^2} + \frac{\partial^2}{\partial r^2} + \frac{1}{r}\frac{\partial}{\partial r})\psi - \frac{C_6}{|\bm{r} - \bm{r}_A|^6}\psi - \frac{1}{4\pi \varepsilon_0}\frac{q^2 \alpha(0)}{|\bm{r}-\bm{r}_A|^4}\psi  = i\hbar \dt \psi \, .
\end{equation}

We define
\begin{equation}
    V(\bm{r}) = - \frac{C_6}{|\bm{r}|^6}- \frac{1}{4\pi \varepsilon_0}\frac{q^2 \alpha(0)}{|\bm{r}|^4}\, ,
\end{equation}
thus equation \ref{eq:schr_long} becomes
\begin{equation}
    \frac{\hbar^2}{2m}(\frac{\partial^2}{\partial z^2} + \frac{\partial^2}{\partial r^2} + \frac{1}{r}\frac{\partial}{\partial r})\psi + V(\bm{r}-\bm{r}_A) \psi = i\hbar \dt \psi \, .
\end{equation}

We divide the box around the helium atom into a uniform grid. We organize the wavefunction into a $n\cdot m$ vector organized such that 
\begin{eqnarray}
    \psi_1 = \psi(r_1, z_1) \, ,\,
    \psi_2 = \psi(r_2, z_1) \, , \,
    ... \, , \, \\
    \psi_n = \psi(r_n, z_1)\, , \,
    \psi_{n+1} = \psi(r_1, z_2) \, , \,
    ... \, , \, \\
    \psi_{n \cdot m -1} = \psi(r_{n-1}, z_m) \, , \, \psi_{n \cdot m} = \psi(r_n, z_m)\, .
\end{eqnarray}
In our case we use n = 256 and m = 256. With r going from 0nm to 8nm and z going from -4nm to 4nm, that makes $\delta r = 7.8$pm and we use a timestep of $\frac{dr}{32v_z}$ with $v_z$ being the velocity of the helium atom. 

The operator $\mathcal{A}_+ = (1 + \frac{i \hat{H}}{2\hbar}\delta t)$ can then be expressed as a $m\cdot n \times m\cdot n$-matrix. The resulting matrix is a block tridiagonal matrix
\begin{equation}
    \mathcal{A}_+ = 
    \begin{bmatrix}
        D_1 & U_1 & 0 & 0 & \dots & 0 & 0 & 0 \\
        L_2 & D_2 & U_2 & 0 & \dots & 0 & 0 & 0 \\
        0 & L_3 & D_3 & U_3 & \dots & 0 & 0 & 0\\
        \vdots & \vdots & \vdots & \vdots & \ddots & \vdots & \vdots & \vdots\\
        0 & 0 & 0 & 0 & \dots & L_{m-1}& D_{m-1} & U_{m-1} \\
        0 & 0 & 0 & 0 & \dots & 0 & L_m &  D_m 
    \end{bmatrix} \, ,
\end{equation}
with
\begin{equation}
    U_i = L_i =  -i\frac{\hbar \delta t}{4m \delta z^2} \mathbf{1}_{n\times n}\, ,
\end{equation}
and 
\begin{equation}
    D_i =
    \begin{bmatrix}
        d_{i,1} & u_1 & 0 & 0 & \dots &  0 & 0 & 0\\
        l_2 & d_{i, 2} & u_2 & 0 & \dots & 0 & 0 & 0\\
        0 & l_3 & d_{i, 3} & u_3 & \dots & 0 & 0 & 0\\
        \vdots & \vdots & \vdots & \vdots & \ddots & \vdots & \vdots & \vdots \\
        0 & 0 & 0 & 0 & \dots & l_{n-1} & d_{i, n-1} & u_{n-1} \\
        0 & 0 & 0 & 0 & \dots & 0 & l_n & d_{i, n}
    \end{bmatrix} \, ,
\end{equation}
where
\begin{eqnarray}
    d_{i, j} = 1 + (\frac{\hbar^2}{m \delta r^2} + \frac{\hbar^2}{m \delta z^2} + V_{i,j})\frac{i\delta t}{2\hbar} \, ,\\
    u_j = -i\frac{\hbar\delta t}{4m\delta r^2} - i\frac{1}{j}\frac{\hbar \delta t}{2m \delta r^2}\, , \\
    l_j = -i\frac{\hbar\delta t}{4m\delta r^2} + i\frac{1}{j}\frac{\hbar \delta t}{2m \delta r^2}\, , \\
    u_1 = 0 \, .
\end{eqnarray}
The operator $\mathcal{A}_-$ is the complex conjugate of $\mathcal{A}_+$.

Performing a time step is therefore equivalent to solving the matrix equation
\begin{equation}
    \mathcal{A}_+ \bm{\psi}_{t+\delta t} = \bm{b}_t ,
\end{equation}
with
\begin{equation}
    \bm{b}_t = \mathcal{A}_- \bm{\psi}_t\, .
\end{equation}

can then be solved by iteration if certain criteria are met\cite{ParallelMatrixAlgorithm}.
The block matrix is solved by iteratively by solving
\begin{equation}
    D_i \bm{x}_i = \bm{b} - L_i \bm{x}_{i-1} - U_i \bm{x}_{i+1}\, ,
\end{equation} 
which should have $\bm{x}$ converge towards a solution\cite{ParallelMatrixAlgorithm}.
Since $D_i$ is a tridiagonal matrix it can be solved quickly using the tridiagonal matrix algorithm \cite{niyogi2006introduction}.

At the end of each time step we multiply the parts inside the van der Waals radius of the atom by 0, that is $|\bm{r} - \bm{r}_A| < r_{\rm vdW}$, and we move the box according to according to its velocity, $\bm{r}_{A, new} = \bm{r}_A -\bm{v} \delta t$, with $\bm{v}$ being the velocity of the helium atom.

We do a total of 6 runs, 3 different velocities for nitrogen and 3 different velocities for boron. The starting shape of the wavepacket is such that the wavepacket sqared is a Gaussian
\begin{equation}
    \psi_0(\bm{r}) = \frac{1}{\sqrt{\sqrt{2\pi^3} \sigma_z \sigma_r^2}} \exp{\left[-\frac{r^2}{4\sigma_r} - \frac{z^2}{4\sigma_z}\right]}\, , 
\end{equation}
with $\sigma_r = \sigma_z = 8$Å. We start the propagation with the center of our box at 60Å before the N or B atom and end it at 40Å past the atoms. The final wavefunction is then compared to the first one to find the amount that passed within the van der Waals radius.

\subsection{Atomic charges in hBN}

\begin{table}[!h]
    \centering
    \caption{Charges (in terms of unit charge $|e|$) around atoms in h-BN.
        \label{tab:charges}
    }
    \begin{tabular}{l|c|c|l}
        \textbf{Material} & \textbf{N Charge} & \textbf{B Charge} & \textbf{Method} \\
        \hline
        hBN pristine\cite{wang_local_2016} & -0.47 & +0.47 & Bader VASP \\
        hBN pristine\cite{yin_triangle_2010} & -0.67 & +0.67 & Bader VASP \\
        hBN pristine\cite{suzuki_designing_2022} & -2.28 & +2.28 & Bader GPAW \\
        hBN pristine\cite{altintas_intercalation_2011} & -0.17 & +0.17 & Hirshfeld Abinit \\
        hBN pristine ours & -2.19 & +2.19 & Bader \\
        hBN pristine ours & -0.20 & +0.20 & Hirshfeld \\
        PQP+\cite{cui_squeezing_2014} & -1.1 & & Bader GPAW \\
        N edge in hBN hole\cite{yin_triangle_2010} & -1.59 & & Bader VASP \\
        B edge in hBN hole\cite{yin_triangle_2010} & & +1.50 & Bader VASP \\\hline
        Circular hole (6 Ang) ours & -0.39 & 0.38 & Hirshfeld \\
        Circular hole (10 Ang) ours & -0.32 & 0.32 & Hirshfeld \\
        Elliptical hole ours & -0.39 & 0.38 & Hirshfeld \\
        Snowflake hole ours & -0.29 & 0.24 & Hirshfeld \\
        \hline
    \end{tabular}
\end{table}
The definition of atomic charges in molecules suffers from a degree of arbitrariness, such that different schemes for the determination have been
devised. Tab. \ref{tab:charges} compares values obtained from Bader 
and Hirshfeld partitioning from the literature and our calculations.
All results show the same qualitative trend in that the nitrogen
atom is negatively charged and the boron atom is positively charged
contrary to chemical understanding. 
Generally the charges determined by the Bader method are larger than these
from Hirshfeld partitioning.

There is a large deviation between Bader results determined from different DFT codes, but the Hirshfeld values are in much better agreement. The origin of the 
difference in the Bader values is beyond the scope of this investigation.
We therefore use the Hirshfeld charges in our manuscript due to their consistency.
Atomic charges near to the holes appear to be larger than these of pristine hBN. 
The maximal Hirshfeld charges we find do not exceed $\pm 0.39 |e|$, however.

\end{document}